\documentclass[preprintnumbers,floats,twocolumn,prd,aps]{revtex4}
\usepackage{amssymb,amsmath,epsfig}

\newcommand{\Lie}{{\pounds}}

\begin{document}

\title{Accretion of Ghost Condensate by Black Holes}
\author{Andrei V. Frolov}\email{afrolov@stanford.edu}
\affiliation{
  KIPAC/SITP, Stanford University\\
  Stanford, CA, 94305-4060
}
\date{April 27, 2004}

\begin{abstract}
  The intent of this letter is to point out that the accretion of a
  ghost condensate by black holes could be extremely efficient. We
  analyze steady-state spherically symmetric flows of the ghost fluid
  in the gravitational field of a Schwarzschild black hole and
  calculate the accretion rate. Unlike minimally coupled scalar field
  or quintessence, the accretion rate is set not by the cosmological
  energy density of the field, but by the energy scale of the ghost
  condensate theory. If hydrodynamical flow is established, it could be
  as high as tenth of a solar mass per second for 10MeV-scale ghost
  condensate accreting onto a stellar-sized black hole, which puts
  serious constraints on the parameters of the ghost condensate model.
\end{abstract}

\preprint{SU-ITP-04-13}
\preprint{SLAC-PUB-10473}
\maketitle

\section{Introduction}

Prompted by an increasingly precise experimental measurements of
cosmological parameters, and in particular detection of acceleration of
the universe due to an unknown source which looks like a cosmological
constant, in the recent years there has been a wide discussion in the
literature about modifications of Einstein gravity on cosmological
scales as a possible alternative to dark matter and/or energy. However,
finding a self-consistent and well-motivated theory which agrees with
all the observations is proving to be quite a challenge.

Recently, Arkani-Hamed {\it et.\ al.}\ proposed a model
\cite{Arkani-Hamed:2003uy}, dubbed {\it a ghost condensation}, which
they argued to be consistent with all experimental observations and
provide an interesting modification of gravity in the infrared, with
potential applications to inflation \cite{Arkani-Hamed:2003uz}, dark
matter and cosmological constant problems. It involves an introduction
of a scalar field which develops a non-zero expectation value of its
(timelike) gradient in vacuum, due to non-trivial kinetic term in the
action. Such modifications of the scalar field kinetic term were
considered earlier on phenomenological grounds in the model known as
$k$-inflation \cite{Armendariz-Picon:1999rj, Garriga:1999vw}.

The ghost condensate model has been studied on perturbative level in
effective field theory \cite{Arkani-Hamed:2003uy}, which already leads
to interesting consequences such as star trails \cite{Dubovsky:2004qe,
Peloso:2004ut}. We look at the ghost condensate from a slightly
different perspective, namely we would like to investigate its
behaviour in the strong gravitational field, for instance, near a
Schwarzschild black hole
\begin{equation}\label{eq:metric}
  ds^2 = - f(r) dt^2 + \frac{dr^2}{f(r)} + r^2 d\Omega_n^2,
\end{equation}
where $f(r) = 1 - r_g/r$, and $r_g = 2Gm$ is a gravitational radius of
Schwarzschild black hole of mass $m$. The problem is similar to
interaction of a cosmological scalar field with a black hole
\cite{Frolov:2002va}, so one would expect ghost condensate to be
accreted by a black hole.

Accretion of fluid onto a black hole has long been an important problem
in astrophysics \cite{Bondi:1952ni}. Spherically symmetric steady-state fluid accretion
onto a Schwarzschild black hole was derived in Ref.~\cite{Michel:1972}.
Minimally coupled scalar field \cite{Jacobson:1999vr} and quintessence
\cite{Frolov:2002va, Bean:2002kx} accrete onto black holes as well,
although the accretion rate is limited by the cosmological density of
the field \cite{Frolov:2002va}. Accretion of exotic matter fields can
lead to unusual results. For instance, accretion of a phantom energy
(which violates energy dominance conditions) {\it decreases} the black
hole size \cite{Babichev:2004yx}.

In this letter, we calculate the steady-state accretion rate of the
ghost condensate by a black hole, and point out that it could be
extremely efficient. This puts serious constraints on the parameters
of the ghost condensate model.

\section{Ghost Condensate as a Fluid}

Ghost condensate model adds a non-minimally coupled scalar field to
Einstein theory of gravity. However, instead of the usual kinetic term
\begin{equation}\label{eq:X}
  X = -(\nabla\phi)^2,
\end{equation}
the action is assumed to involve a more complicated function of the
field gradient squared
\begin{equation}\label{eq:action}
  S =  \int \left\{\frac{R}{16\pi G} + M^4 P(X)\right\}\, \sqrt{-g}\, d^4 x,
\end{equation}
as well as higher-derivative terms. We will ignore higher-derivative
terms in what follows. They complicate calculations significantly,
and we are concerned with large scale flows, while one would expect
higher-derivative terms to be important on short scales.

\begin{figure}
  \epsfig{file=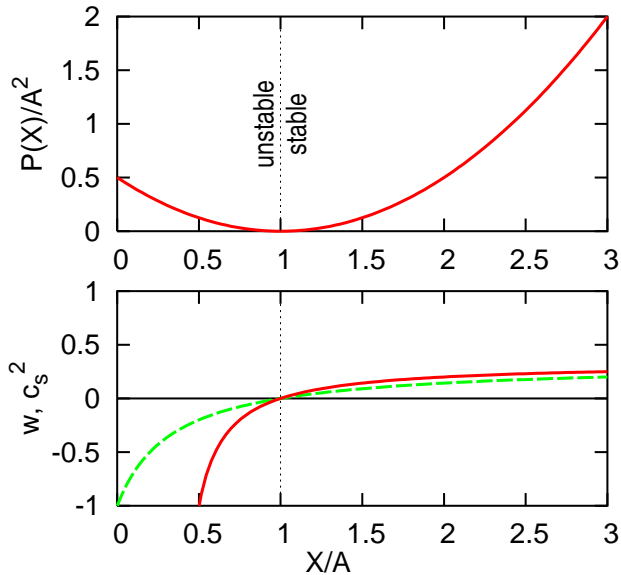, width=246pt}
  \caption{
    Ghost condensate kinetic term (top) and equivalent fluid description
    (bottom). Equation of state $w$ and sound speed $c_s^2$ are shown
    by dashed and solid curves, respectively.
  }
  \label{fig:ghost}
\end{figure}

As the ghost field $\phi$ is not directly coupled to other fields, its
dimensionality and normalization are arbitrary. If the field $\phi$ is
chosen to have dimension of length, the field gradient $X$ and the
function $P$ are dimensionless, and the only dimensional quantity in
the ghost sector is $M$, which sets the overall energy scale of the
ghost condensate. The specific form of the function $P(X)$ is not
rigidly fixed, although the defining feature of the ghost condensate
model is that $P$ has a minimum at non-vanishing (timelike) value of
the field gradient. Because of that, the ghost field rolls even in its
vacuum state. The simplest choice for $P$ with this property is
\begin{equation}\label{eq:P}
  P(X) = \frac{1}{2}\, (X-A)^2,
\end{equation}
illustrated in Fig.~\ref{fig:ghost}. One could also add a cosmological
constant term $\Lambda$, but since it is not accreted by a black hole,
we will not discuss it further.

Variation of the action (\ref{eq:action}) with respect to the ghost
field $\phi$ yields equation of motion
\begin{equation}\label{eq:eom}
  \nabla_\mu\left[P'(X) \nabla^\mu\phi\right] \equiv
  \frac{1}{\sqrt{-g}}\, \partial_\mu\left[\sqrt{-g}\, P'(X)\, \partial^\mu\phi\right] = 0.
\end{equation}
The equation of motion is implied by conservation of the
stress-energy tensor, which for the ghost condensate is
\begin{equation}\label{eq:T:ghost}
  T_{\mu\nu} = 2 M^4 P'(X) \phi_{;\mu}\phi_{;\nu} + M^4 P(X) g_{\mu\nu}.
\end{equation}
Configurations with $P'(X) \equiv 0$ solve the equation of motion
identically for any spacetime metric. However, such configurations are
indistinguishable gravitationally from a purely Einstein theory, as the
stress-energy tensor becomes trivial as well.

The stress-energy tensor of the ghost condensate (\ref{eq:T:ghost})
can be transformed into that of a perfect fluid
\begin{equation}\label{eq:T:fluid}
  T_{\mu\nu} = (\rho+p) u_\mu u_\nu + p g_{\mu\nu}
\end{equation}
by a formal identification
\begin{equation}\label{eq:fluid}
  \rho = M^4 (2XP'-P), \hspace{1em}
  p = M^4 P, \hspace{1em}
  u_\mu = \frac{\phi_{;\mu}}{\sqrt{X}}.
\end{equation}
The fluid analogy is very useful in understanding the physics behind
the solutions of the ghost equation of motion (\ref{eq:eom}), although
it is not an exact correspondence. Unlike ordinary fluids, ghost
condensate is {\em irrotational}, that is, the vorticity tensor of the
flow $u^\mu$ vanishes identically
\begin{equation}\label{eq:vorticity}
  \omega_{\alpha\beta} = \frac{1}{2} \left(
    u_{\alpha;\mu}q^\mu_{\ \beta} - u_{\beta;\mu}q^\mu_{\ \alpha}
  \right) \equiv 0,
\end{equation}
where $q_{\mu\nu} = g_{\mu\nu} + u_\mu u_\nu$. This is a direct
consequence of the vector flow $u^\mu$ being derived from a scalar.

Important parameters of the fluid are its equation of state and sound speed
\begin{equation}
  w \equiv \frac{p}{\rho}, \hspace{1em}
  c_s^2 \equiv \frac{dp}{d\rho} = \frac{p'}{\rho'}.
\end{equation}
For the ghost condensate with kinetic term (\ref{eq:P}), they are
\begin{equation}
  w = \frac{X-A}{3X+A}, \hspace{1em}
  c_s^2 = \frac{X-A}{3X-A}.
\end{equation}
The equation of state and the sound speed change from dust-like in the
minimum $X=A$ to radiation-like for large displacements $X \gg A$, as
shown in Fig.~\ref{fig:ghost}. Configurations with $X<A$ are unstable,
as the sound speed squared becomes negative.

Cosmological expansion of the universe dilutes the density of the
homogeneous ghost condensate and drives its gradient toward the minimum
of the kinetic term $P$ \cite{Arkani-Hamed:2003uy}. Dust-like equation
of state of the ghost condensate near its minimum lends itself to
interpretation of ghost condensate as a dark matter in late-time
cosmology \cite{Arkani-Hamed:2003uy}. If this is the case, the energy
density of the ghost condensate $\rho$ is the energy density of the
dark matter, and displacement of the ghost condensate from the minimum
is small, $P' = \rho/(2M^4A)$, but non-zero. ``Modification of
gravity'' depends on the excitations in the ghost condensate, which
displace the ghost condensate from its minimum and carry energy density
\cite{Arkani-Hamed:2003uy}. Indeed, as the ghost condensate with $P'
\equiv 0$ is {\em identical} to cosmological constant, such restriction
would greatly diminish the attractiveness of the ghost condensate
model. In view of the above, we assume the ghost condensate distribution
which is homogeneous far from the black hole, but could be displaced
from the minimum (by a small amount).

\section{Steady-State Accretion}

\begin{figure*}
  \epsfig{file=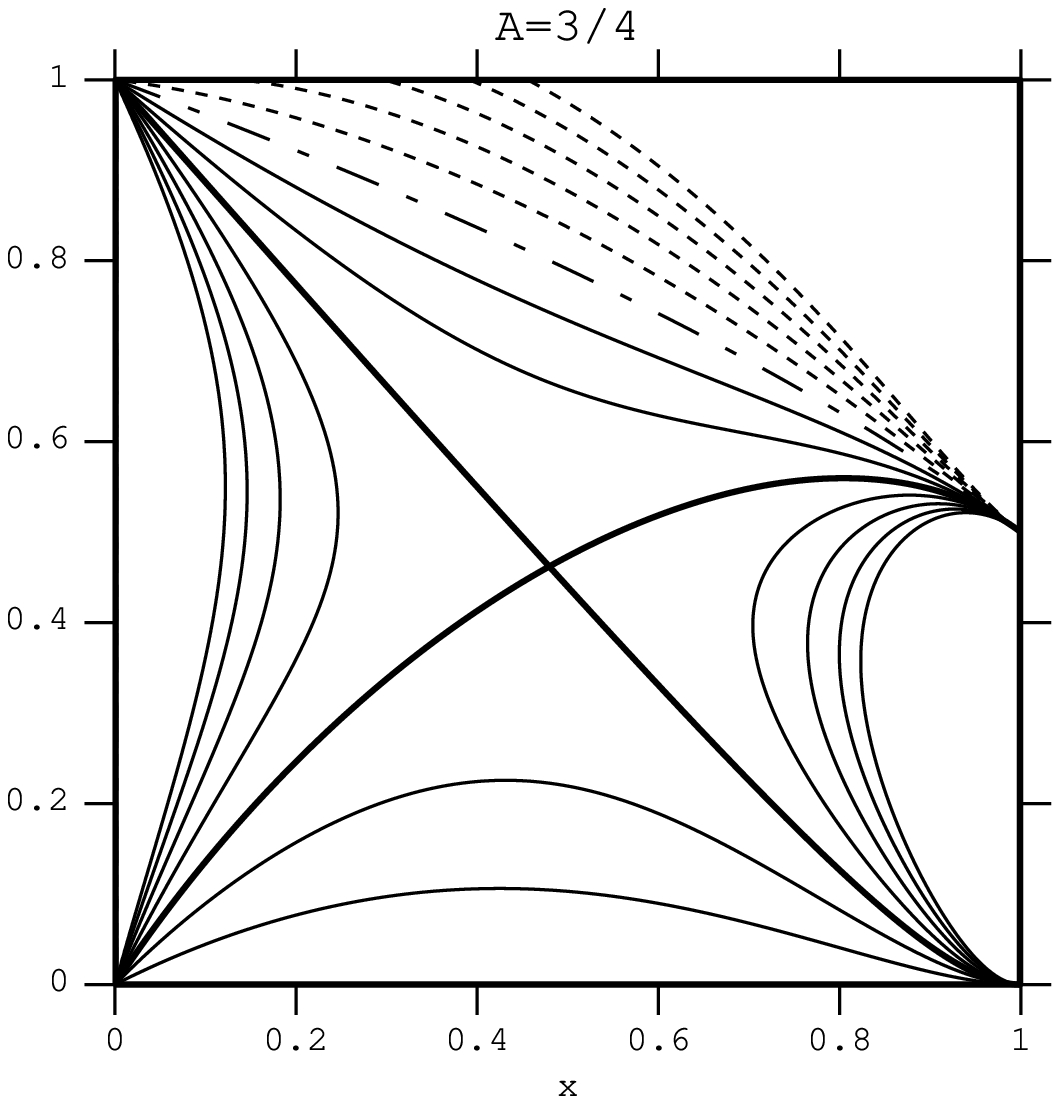, width=246pt}
  \epsfig{file=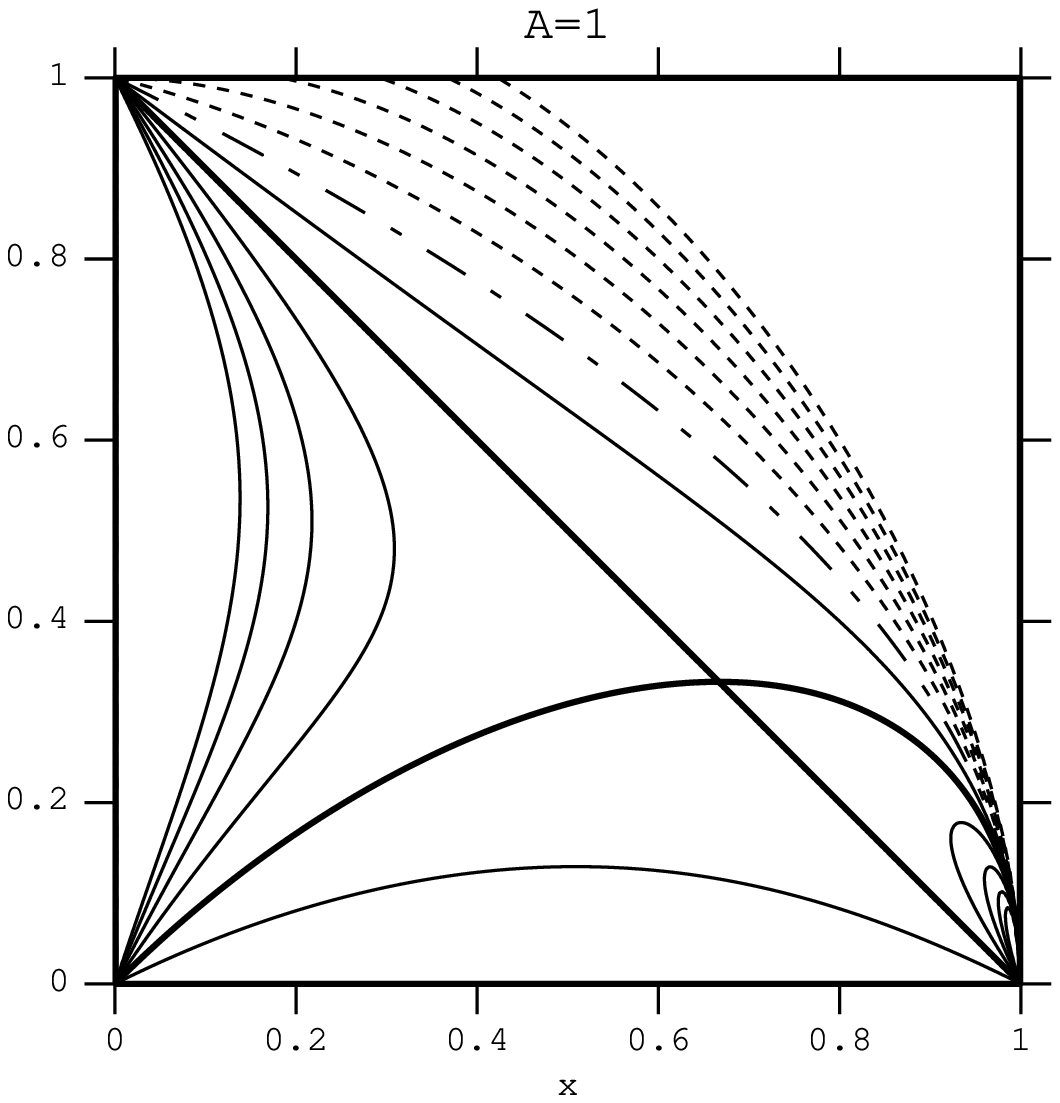, width=246pt}
  \vspace{-1em}
  \caption{
    Flow diagrams $v(x)$ of the ghost condensate accretion onto a black
    hole for $A=3/4$ (left) and $A=1$ (right). Inflow, no-flow, and
    (unstable) outflow branches are shown by solid, dot-dash, and
    dotted lines correspondingly. Flow trajectories passing through
    the critical point are emphasized by thicker lines. Negative $v$
    region corresponds to reversed flow direction, and is not shown.
  }
  \label{fig:flow}
\end{figure*}

Steady-state accretion means that the flow of the field (i.e.\ its
gradient) does not change with time, that is
$\Lie_{\partial_t}(\nabla_\mu \phi) = \partial_\mu \partial_t \phi = 0$,
which in turn implies that $\partial_t \phi$ is constant (and can be set
to one by a choice of the field normalization). Therefore, a general
steady-state spherically symmetric field configuration is of the form
\begin{equation}\label{eq:ss}
  \phi = t + \psi(r),
\end{equation}
and, in Schwarzschild spacetime (\ref{eq:metric}), has a gradient
\begin{equation}\label{eq:ss:X}
  X = \frac{1-(\partial^*_r \psi)^2}{f(r)},
\end{equation}
where we introduced a ``tortoise'' derivative $\partial^*_r \equiv f(r) \partial_r$.

For steady-state accretion of the spherically symmetric ghost
condensate profile (\ref{eq:ss}) onto a Schwarzschild black hole
(\ref{eq:metric}), the equation of motion (\ref{eq:eom}) becomes
\begin{equation}\label{eq:ss:eom}
  \partial^*_r(r^2 P' \partial^*_r \psi) = 0,
\end{equation}
which can be immediately integrated to yield the flow conservation equation
\begin{equation}\label{eq:flow}
  P' \partial^*_r \psi = \alpha\, \frac{r_g^2}{r^2}.
\end{equation}
The meaning of the dimensionless constant of integration $\alpha$
becomes clear if one looks at the accretion rate
\begin{equation}\label{eq:rate}
  \dot{m} = 4\pi r^2 T^r_t = 2\alpha\cdot 4\pi r_g^2 M^4,
\end{equation}
which does not depend on $r$ and describes a steady-state transfer of
mass from infinity into a black hole. The numerical value of the
coefficient $\alpha$ is picked by the solution of the flow equation
(\ref{eq:flow}) that is regular at the horizon and becomes homogeneous
far from the black hole.

The flow equation (\ref{eq:flow}) is actually algebraic in
$\partial^*_r \psi$, and could be analyzed for an arbitrary function
$P$. We will restrict our discussion to the ghost condensate with
kinetic term (\ref{eq:P}) and further assume $A \le 1$, as the choice
$A>1$ places the solution (\ref{eq:ss}) on the unstable branch of the
kinetic term far from a black hole and is not physically relevant.
Introducing the short-hand notation $v \equiv \partial^*_r \psi$ and
$x \equiv f(r)$, the flow equation (\ref{eq:flow}) can be written as
\begin{equation}\label{eq:alpha}
  \left(\frac{1-v^2}{x} - A\right) \frac{v}{(1-x)^2} = \alpha.
\end{equation}
Solutions $v(x)$ for various values of $\alpha$ are shown in
Fig.~\ref{fig:flow}. Although cubic equation (\ref{eq:alpha}) can be
directly solved in radicals, the flow is more readily analyzed using
standard phase space diagram techniques.

\begin{figure}
  \epsfig{file=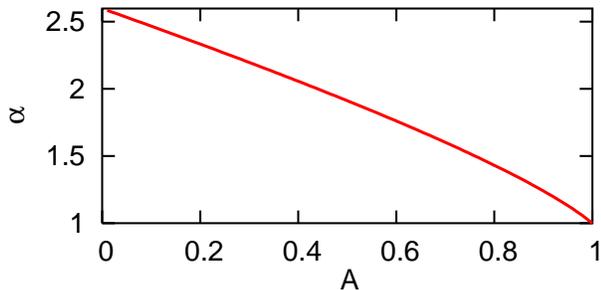, width=246pt}
  \caption{
    Dependence of the accretion rate coefficient $\alpha$ on $A$.
  }
  \label{fig:alpha}
\end{figure}

Both at the horizon ($x=0$) and infinity ($x=1$), all flow trajectories
converge to one of three roots
\begin{equation}\label{eq:roots}
  \begin{array}{r@{\hspace{1em}}l}
    x = 0: & v_0 = 0, \pm 1\\
    x = 1: & v_1 = 0, \pm \sqrt{1-A}\\
  \end{array}.
\end{equation}
All flow trajectories must start and end at these roots, and they do
not intersect except at the critical points. The critical points are
defined as the points where the full differential of (\ref{eq:alpha}),
\begin{equation}
  - \frac{3v^2-1+Ax}{x(1-x)^2}\, dv - \frac{(1-v^2)(1-3x) + 2Ax^2}{x^2(1-x)^3}\, v\, dx = 0,
\end{equation}
becomes degenerate, i.e.\ when coefficients in front of $dv$ and $dx$
both vanish. In the positive $v$ region, there is (at most) one critical
point
\begin{equation}\label{eq:critical}
  v_*^2 = \frac{A + \sqrt{A^2-36A+36}}{18}, \hspace{1em}
  x_* = \frac{1-3v_*^2}{A}.
\end{equation}
Regularity at the horizon for ingoing flow demands that $v_0=1$, while
proper fall-off at infinity requires $v_1=0$. For $A<1$, the only flow
trajectory that connects the two is the one that passes through the
critical point (\ref{eq:critical}), as it is clear from the left panel
of Fig.~\ref{fig:flow}. The flow starts out subsonic at infinity, and
turns supersonic at the critical point. Thus, the accretion rate for
the steady-state flow is set by the local conditions at the supersonic
transition, which happens in the immediate vicinity of the black hole,
and depends little on the actual boundary conditions at infinity.
The coefficient $\alpha$ is calculated by evaluating equation
(\ref{eq:alpha}) at the critical point (\ref{eq:critical}). The
resulting expression is straightforward, but cumbersome for arbitrary
$A$, so we will not write it down here. Instead, we will show the graph
of $\alpha$ as a function of $A$ in Fig.~\ref{fig:alpha}. The
coefficient $\alpha$ decreases monotonically from $3\sqrt{3}/2$ at
$A=0$ to $1$ at $A=1$. Note that it does not vanish even as
displacement of the ghost condensate from the minimum approaches zero.

Asymptotics of the ghost condensate profile far from the black hole are
easy to analyze directly from the flow equation (\ref{eq:alpha}). For
the slow flow ($v \ll 1$), $v^2$ term in equation (\ref{eq:alpha}) can
be neglected, and we have
\begin{equation}\label{eq:infty}
  v \simeq \alpha\, \frac{(1-x)^2}{x^{-1}-A}, \hspace{1em}
  \psi \simeq \frac{\alpha }{A}\, r_g \ln\frac{r}{A r_g + (1-A)r}.
\end{equation}
As can be seen from the above expressions, sufficiently far from the
black hole the field gradient $v$ falls off at infinity as $r^{-2}$,
while the field profile levels off as $r^{-1}$. The field profile is
influenced by the black hole and deviates from homogeneous significantly
inside a ``sphere of influence'' of radius $r_i = A r_g/(1-A)$.

The case of $A=1$ is special, and is shown on the right panel of
Fig.~\ref{fig:flow}. The three roots at infinity merge into one triple
root at $v_1=0$, and one can get from infinity to horizon without going
through a critical point. These solutions correspond to a dust-like
flow with $0\le\alpha<1$, and are always supersonic. However, their
gradient $v$ falls off at infinity only as $r^{-\frac{1}{2}}$, which
means that the ghost field does not become homogeneous far from a black
hole, but in fact grows as $r^{\frac{1}{2}}$. In particular, the
trivial solution ($P' \equiv 0$, $\alpha=0$) is
\begin{equation}
  \phi = t + 2r_g^{\frac{1}{2}} \left[r^{\frac{1}{2}} - r_g^{\frac{1}{2}} \text{arctanh} \sqrt{\frac{r_g}{r}}\, \right].
\end{equation}
The likely reason behind the change in field asymptotic is that
spherically symmetric dust accretion is not steady-state. The
accretion rate is ever growing, as the dust from larger and larger
volume falls inside the black hole.

For $A=1$, the flow trajectory which passes through a critical point
($\alpha=1$) is simply $v = 1-x = r_g/r$. The corresponding field
profile is
\begin{equation}
  \phi = t + r_g \ln\left[\frac{r}{r_g} - 1\right].
\end{equation}
Its asymptotic at infinity is also non-homogeneous, but the growth is
only logarithmic. This solution emerges from steady-state flow
solutions (\ref{eq:infty}) with $A<1$ in the limit when the sphere of
influence of the black hole grows infinitely large.

\section{Discussion}

In the last section, we calculated the steady-state accretion rate of
the ghost condensate by a black hole for spherically symmetric flows.
The most important result of the calculation is that the dimensionless
coefficient $\alpha$, which determines the accretion rate of the flow,
is bounded below by $1$ even as density of the ghost condensate far
from a black hole becomes vanishingly small. This means that it is the
energy scale $M$ of the ghost condensate theory that sets the accretion
rate, and not the cosmological abundance of the ghost condensate field
as one might have naively expected.

The physical reason for this is that unlike the usual dark matter, the
homogeneous ghost condensate does not have angular momentum and can be
prevented from falling onto a gravitating body only by building up
pressure support, which requires significant displacement from the
minimum and densities of order $M^4$. This consideration alone might
pose significant problems for ghost condensate as a realistic dark
matter candidate, as dark matter galaxy halos are thought to be {\em
virialized} and not pressure-supported. This seems impossible to
achieve in the framework of the ghost condensate model, unless the
condensate becomes strongly non-homogeneous and breaks up into
``particles'', in which case it is hard to view it in any sense as a
modification of gravity.

Up to a numerical coefficient of order one, the steady-state accretion
rate is equal to the energy density $M^4$ falling down through the
horizon area $4\pi r_g^2$ at the speed of light. The top value of 10MeV
for the ghost energy scale quoted in \cite{Arkani-Hamed:2003uy}
corresponds to a rather high density
\begin{equation}
  (10\text{MeV})^4 =
  \frac{(10\text{MV} \cdot e)^4}{\hbar^3 c^5} =
  2.32 \cdot 10^{12} \frac{\text{kg}}{\text{m}^3}.
\end{equation}
If the steady-state flow of the kind we considered is established, the
accretion rate of a 10MeV-scale ghost condensate by an astrophysical
black hole would be enormous
\begin{equation}
  \dot{m} = 0.08\,\alpha\, \frac{M_\odot}{\text{s}}
    \left(\frac{r_g}{3\text{km}}\right)^2
    \left(\frac{M}{10\text{MeV}}\right)^4.
\end{equation}
To avoid rapid black hole growth and its astrophysical consequences,
energy scale $M$ of the ghost condensate should be significantly less
than 10MeV. Stellar-size black hole would double in size over the
lifetime of the universe (roughly 14Gyrs, or $4\cdot10^{17}$s) for the
ghost energy scale of order 1keV. This estimate goes down to 10eV for
super-massive ($10^9 M_\odot$) black holes.

In our calculations, we ignored the backreaction of the accreting ghost
condensate matter onto a black hole metric. This is a good approximation
for low accretion rates and ghost energy scales. However, if the
accretion rate becomes as large as the above estimate, backreaction can
no longer be ignored, both because the density of the ghost condensate
near black hole is high and a large supply of ghost matter far from the
black hole is needed to sustain the flow. Although proper treatment of
backreaction is unlikely to alleviate the problem of excessive
accretion rates, as it is caused by the largeness of the effect in the
first place, sustainability of the steady-state flow with such enormous
accretion rates is questionable. It would seem more likely that the
black holes would completely evacuate all ghost matter from their
gravitational potential wells, growing in the process.

Spherically symmetric steady-state accretion is an idealized situation,
of course. We have not considered time-scale required to establish such
flows, the role of initial conditions, motion of the black hole with
respect to the condensate, or what happens if the ghost field becomes
highly inhomogeneous. All of these are much harder problems, and it
might turn out that some factors prevent the accretion from settling
into an efficient steady-state regime. Still, having ghost condensate
capable of such high accretion rates is alarming, and the issue should
be further addressed by the ghost condensate scenario.

\section*{Acknowledgments}

I am grateful to Shinji Mukohyama for sharing his insights into the
ghost condensate dynamics. I would like to thank Renata Kallosh, Lev
Kofman, Andrei Linde, and Marco Peloso for their helpful comments. This
work was supported by Kavli Institute for Particle Astrophysics and
Cosmology and Stanford Institute for Theoretical Physics.



\begin{thebibliography}{99}

\bibitem{Arkani-Hamed:2003uy}
N.~Arkani-Hamed, H.~C.~Cheng, M.~A.~Luty and S.~Mukohyama,
{\it Ghost condensation and a consistent infrared modification of gravity},
{\tt hep-th/0312099}.

\bibitem{Arkani-Hamed:2003uz}
N.~Arkani-Hamed, P.~Creminelli, S.~Mukohyama and M.~Zaldarriaga,
{\it Ghost inflation},
JCAP {\bf 0404}, 001 (2004)
[{\tt hep-th/0312100}].

\bibitem{Armendariz-Picon:1999rj}
C.~Armendariz-Picon, T.~Damour and V.~Mukhanov,
{\it $k$-inflation},
Phys.\ Lett.\ B {\bf 458}, 209 (1999)
[{\tt hep-th/9904075}].

\bibitem{Garriga:1999vw}
J.~Garriga and V.~F.~Mukhanov,
{\it Perturbations in $k$-inflation},
Phys.\ Lett.\ B {\bf 458}, 219 (1999)
[{\tt hep-th/9904176}].

\bibitem{Dubovsky:2004qe}
S.~L.~Dubovsky,
{\it Star tracks in the ghost condensate},
{\tt hep-ph/0403308}.

\bibitem{Peloso:2004ut}
M.~Peloso and L.~Sorbo,
{\it Moving sources in a ghost condensate},
{\tt hep-th/0404005}.

\bibitem{Frolov:2002va}
A.~Frolov and L.~Kofman,
{\it Inflation and de Sitter thermodynamics},
JCAP {\bf 0305}, 009 (2003)
[{\tt hep-th/0212327}].

\bibitem{Bondi:1952ni}
H.~Bondi,
{\it On spherically symmetrical accretion},
Mon.\ Not.\ Roy.\ Astron.\ Soc.\  {\bf 112}, 195 (1952).

\bibitem{Michel:1972}
F.~C.~Michel,
{\it Accretion of matter by condensed objects},
Astrophysics and Space Science {\bf 15}, 153 (1972).

\bibitem{Jacobson:1999vr}
T.~Jacobson,
{\it Primordial black hole evolution in tensor scalar cosmology},
Phys.\ Rev.\ Lett.\  {\bf 83}, 2699 (1999)
[{\tt astro-ph/9905303}].

\bibitem{Bean:2002kx}
R.~Bean and J.~Magueijo,
{\it Could supermassive black holes be quintessential primordial black holes?},
Phys.\ Rev.\ D {\bf 66}, 063505 (2002)
[{\tt astro-ph/0204486}].

\bibitem{Babichev:2004yx}
E.~Babichev, V.~Dokuchaev and Y.~Eroshenko,
{\it Black hole mass decreasing due to phantom energy accretion},
{\tt gr-qc/0402089}.


\end{thebibliography}

\end{document}